\documentclass[11pt]{article}
\usepackage{amsmath}
\usepackage{amssymb}
\usepackage[dvips]{graphicx}
\begin{document}
\title{Analytical potential-density pairs for bars}
\author{D. Vogt\thanks{e-mail: dvogt@ime.unicamp.br} 
\and
P. S. Letelier\thanks{e-mail: letelier@ime.unicamp.br}\\
Departamento de Matem\'{a}tica Aplicada-IMECC, Universidade \\
Estadual de Campinas 13083-970 Campinas, S.\ P., Brazil}
\maketitle
\begin{abstract}
An identity that relates multipolar solutions of the Einstein equations to Newtonian potentials 
of bars with linear densities proportional to Legendre polynomials is used to construct 
analytical potential-density pairs of infinitesimally thin bars with a given linear density profile. 
By means of a suitable transformation, softened bars that are free of singularities are also 
obtained. As an application we study the equilibrium points and stability for the motion 
of test particles in the gravitational field for three models of rotating bars.

\end{abstract}

\section{Introduction}
Bars are a common self-gravitating structure present in disc galaxies. About 50\% of such galaxies
are stongly or weakly barred, including our Milky Way \cite{sw93,bm98}. The effect
of a weak bar is usually represented as a potential in cylindrical coordinates in the form $\Phi(R,\varphi)=\Phi(R)\cos (2\varphi)$ 
 \cite{bt08}. In the case of strong bars, the only exact, self-consistent 
models known are those of Freeman \cite{fr66}, although they present some unrealistic features for
bars. In studies of orbits involving strong bars, they are often modelled as homogeneous 
ellipsoids \cite{d65,m75} or inhomogeneous prolate spheroids \cite{vf72,abmp83,pp83,pf84}. 
These mass distributions have a finite extent. Long \& Murali \cite{lm92} discuss a simple method
to generate analytical potential-density pairs for barred systems that extend to all space.

In this paper we construct analytical potential-density pairs for infinitesimally thin 
and  `softened' bars \cite{lm92} that can be expressed solely in terms of elementary 
functions. The starting point is an identity that relates multipolar solutions of 
the Einstein equations to Newtonian potentials of bars with densities proportional  
to Legendre polynomials. These bars can then be superposed to generate other
bars with a desired density profile. We also use the method of \cite{lm92} to
soften the infinitesimally thin bars. This is presented in Section \ref{sec_bars}.
In Section \ref{sec_eq}  the potentials for barred systems are used to study an
aspect of the motion of test particles in uniform rotating bars, namely  the equilibrium
points (Lagrange points) and their stability. We relate the properties of the
equilibrium points to the mass distribution of the bar models. Section \ref{sec_disc} 
is devoted to the discussion of the results.

\section{Bars with variable densities} \label{sec_bars}

In this section an identity derived by Letelier \cite{l99} will be used as a starting point to construct 
potential-density pairs of bars with various linear density profiles. 
The Newtonian potential of a bar of length 
$2L$ with linear density $\lambda (z)$ located symmetrically along the $z$-axis
is 
\begin{equation} \label{eq_phi1}
\Phi=-G \int_{-L}^{L} \frac{\lambda(z^{\prime})\mathrm{d}z^{\prime}}{\sqrt{R^2+
\left( z-z^{\prime} \right)^2}} \mbox{,} 
\end{equation}
where $G$ is the gravitational constant. Letelier \cite{l99} found the following identity: 
\begin{equation} \label{eq_ide} 
Q_n(\xi)P_n(\eta)=\frac{1}{2} \int_{-L}^{L} \frac{P_n(z^{\prime}/L)\mathrm{d}z^{\prime}}{\sqrt{R^2+
\left( z-z^{\prime} \right)^2}} \mbox{,} 
\end{equation} 
where $P_n$ and $Q_n$ are, respectively, the Legendre polynomials and the Legendre 
functions of the second kind  and $(\xi,\eta)$ are the spheroidal coordinates related to 
the cylindrical coordinates $(R,z)$ through 
\begin{gather}
\xi =(R_1+R_2)/(2L) \text{,} \qquad  \eta = (R_1-R_2)/(2L) \mbox{,} \\
R_1 =\sqrt{R^2+\left(z+L \right)^2} \text{,} \qquad  R_2 = \sqrt{R^2+\left(z-L \right)^2} \mbox{,} 
\end{gather}
with $\xi \geq 1$ and $-1 \leq \eta \leq 1$. Comparing equations (\ref{eq_phi1}) and 
(\ref{eq_ide}), and introducing the mass $M$,  
we see that relation (\ref{eq_ide}) represents a family of bars with linear density
\begin{equation} \label{eq_lambda}
\lambda_n(z)=\frac{M}{2L}P_n(z/L) \mbox{,} 
\end{equation}
associated with a potential $\Phi_n=-GMQ_n(\xi)P_n(\eta)/L$. 

Since the Legendre polynomials form a complete set of functions, the 
members of the family (\ref{eq_lambda}) can be superposed to generate potential-density pairs for bars
with a prescribed density distribution. The simplest case is the bar with constant density  
\begin{equation} \label{eq_lambda0}
\lambda_0= \frac{M}{2L} \mbox{,} 
\end{equation}
whose potential can be expressed in cylindrical coordinates as 
\begin{equation} \label{eq_phi0}
\Phi_0=\frac{GM}{2L} \ln \left[ \frac{z-L+\sqrt{R^2+\left( z-L \right)^2}}
{z+L+\sqrt{R^2+\left( z+L \right)^2}} \right] \mbox{.} 
\end{equation}
To obtain the simple form of equation (\ref{eq_phi0}) from the Legendre function $Q_0$ 
we used the auxiliar functions $\mu_1=z+L+R_1$, $\mu_2=z-L+R_2$ and the  
identities \cite{ll94} 
\begin{gather}
R_1 =\frac{R^2+\mu_1^2}{2\mu_1} \text{,} \qquad  R_2 =\frac{R^2+\mu_2^2}{2\mu_2} \mbox{,} \\
z+L =\frac{\mu_1^2-R^2}{2\mu_1} \text{,} \qquad  z-L =\frac{\mu_2^2-R^2}{2\mu_2} \mbox{.}
\end{gather} 
A bar with maximum of density at the centre and vanishing density at both ends can be 
obtained by the superposition 
\begin{equation} \label{eq_lambda02}
\lambda_{02} =\lambda_0-\lambda_2=\frac{3M}{4L} \left( 1-\frac{z^2}{L^2}\right) \mbox{.} 
\end{equation}
The corresponding potential reads 
\begin{multline}  \label{eq_phi02}
\Phi_{02}=\frac{3GM}{8L^3} \left( R^2+2L^2-2z^2\right) \ln \left[ \frac{z-L+\sqrt{R^2+\left( z-L \right)^2}}
{z+L+\sqrt{R^2+\left( z+L \right)^2}} \right] \\
+ \frac{3GM}{8L^3} \left[ \left( L-3z \right) 
\sqrt{R^2+\left( z+L \right)^2} +\left( L+3z \right) \sqrt{R^2+\left( z-L \right)^2}\right] \mbox{.}
\end{multline}
We will also consider another bar with density obtained by the superposition 
\begin{equation} \label{eq_lambda024} 
\lambda_{024} =\lambda_0+\frac{5}{7} \lambda_2-\frac{12}{7} \lambda_4 =
\frac{15Mz^2}{4L^3} \left( 1-\frac{z^2}{L^2}\right) \mbox{.} 
\end{equation}
The density (\ref{eq_lambda024}) vanishes at the centre and at both ends of the 
bar, and has maxima at $z/L=\pm \sqrt{2}/2$. The associated potential can be 
expressed as 
\begin{multline} \label{eq_phi024}
\Phi_{024}=\frac{15GM}{32L^5}\left( -3R^4+24R^2z^2-4R^2L^2+8z^2L^2-8z^4 \right) \\
\times \ln \left[ \frac{z-L+\sqrt{R^2+\left( z-L \right)^2}}
{z+L+\sqrt{R^2+\left( z+L \right)^2}} \right] + \frac{5GM}{32L^5} \left[ \left(
-55R^2z-9R^2L+26z^2L \right. \right. \\
\left. \left. -22zL^2+50z^3-6L^3 \right) \sqrt{R^2+\left( z-L \right)^2} \right. \\ 
\left. +\left( 55R^2z-9R^2L+26z^2L+22zL^2-50z^3-6L^3 \right) 
\sqrt{R^2+\left( z+L \right)^2}\right] \mbox{.}
\end{multline}

\begin{figure}
\centering
\includegraphics[scale=0.7]{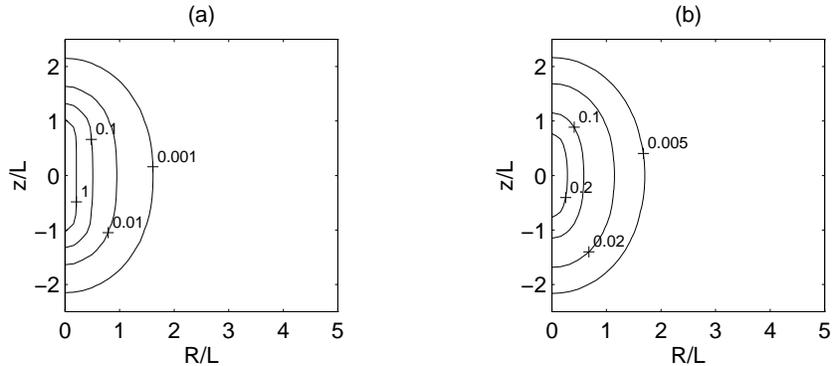}
\caption{Isodensity contours of the 
dimensionless density $\bar{\rho}_0=\rho_0/(M/L^3)$, equation (\ref{eq_rho0}), 
as functions of $R/L$ and $z/L$ for  (a) $b/L=0.25$ and (b) $b/L=0.75$.} \label{fig1}
\end{figure}

\begin{figure}
\centering
\includegraphics[scale=0.7]{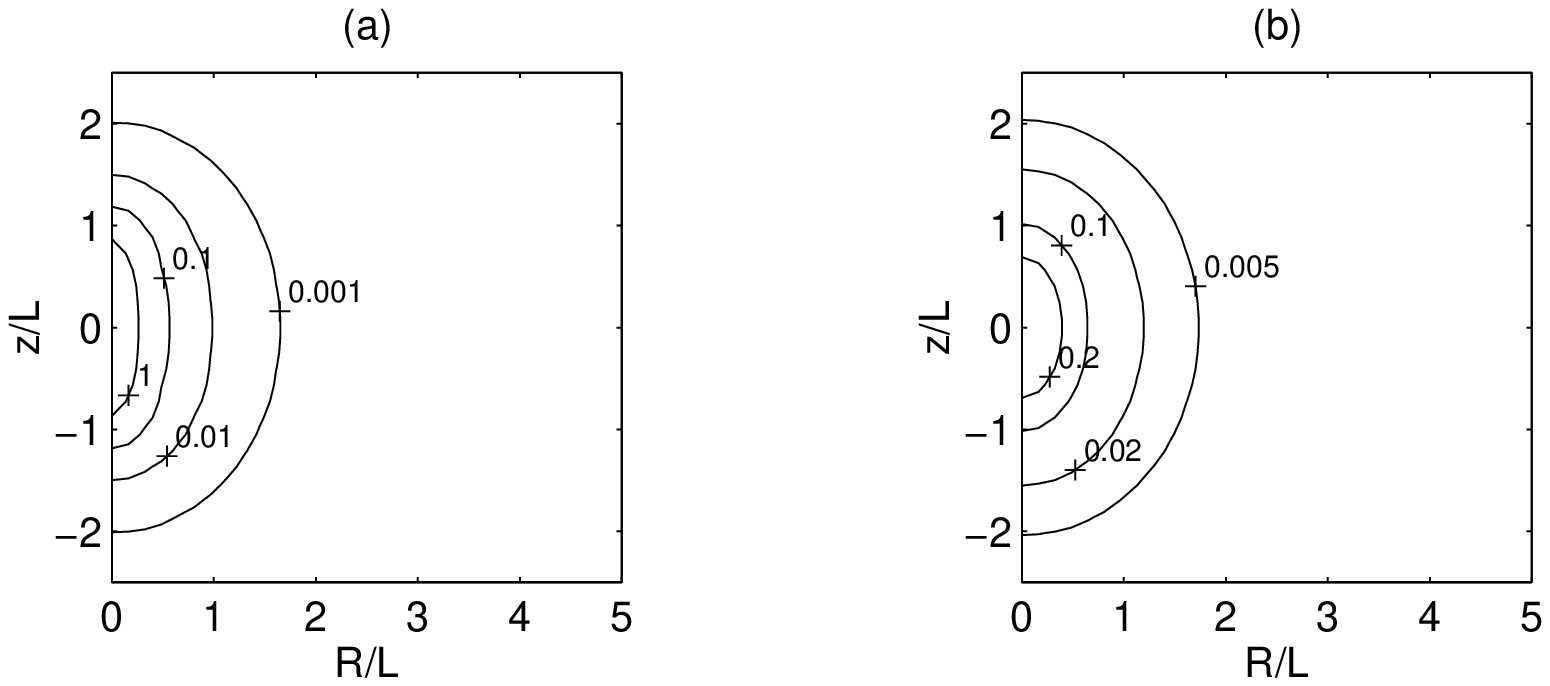}
\caption{Isodensity contours of the 
dimensionless density $\bar{\rho}_{02}=\rho_{02}/(M/L^3)$, equation (\ref{eq_rho02}), 
as functions of $R/L$ and $z/L$ for  (a) $b/L=0.25$ and (b) $b/L=0.75$.} \label{fig2}
\end{figure}

\begin{figure}
\centering
\includegraphics[scale=0.7]{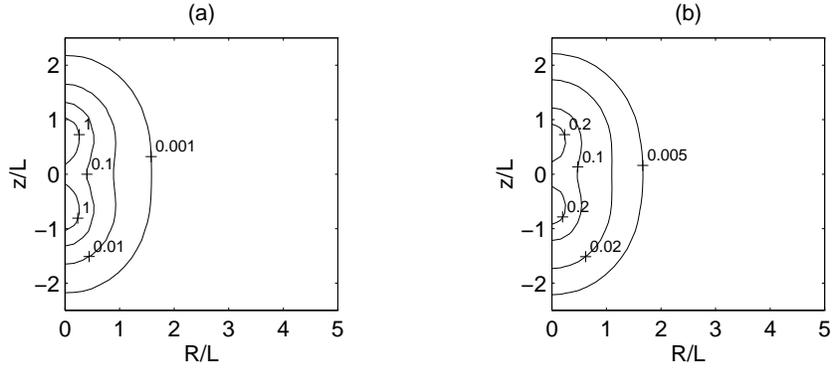}
\caption{Isodensity contours of the 
dimensionless density $\bar{\rho}_{024}=\rho_{024}/(M/L^3)$, equation (\ref{eq_rho024}), 
as functions of $R/L$ and $z/L$ for  (a) $b/L=0.25$ and (b) $b/L=0.75$.} \label{fig3}
\end{figure}

The above potential-density pairs refer to infinitesimally thin bars, thus the
potential is singular along the bar. For astrophysical applications (e.g. galactic
bars) more realistic potentials should be free of singularities. A very simple way to
`soften' these potentials is by making a Plummer-like transformation
$R^2 \rightarrow R^2+b^2$, where $b$ is a non-negative parameter \cite{lm92}.
With this procedure one obtains potential
density-pairs that make a transition between infinitesimally thin bars $(b=0)$ 
and a Plummer sphere $(b \gg L)$ \cite{bt08}. Applying this transformation on
the potentials (\ref{eq_phi0}), (\ref{eq_phi02}) and (\ref{eq_phi024}), the 
corresponding mass density distributions are calculated directly from Poisson 
equation in cylindrical coordinates, 
\begin{equation}
\Phi_{,RR}+\frac{\Phi_{,R}}{R}+\Phi_{,zz}=4\pi G \rho \mbox{.}
\end{equation}
The explicit expressions are given in Appendix \ref{ap_A}. The three mass 
densities are free from singularities and non-negative everywhere. For 
large values of $R$ and $z$, the mass densities decay with $(R^2+z^2)^{-5/2}$,
as can be verified by an asymptotic expansion or simply
by noting that in this limit the densities approach that of the Plummer sphere,
which decays as $(R^2+z^2)^{-5/2}$. Thus, in principle, they 
fall fast enough to put a clear cut-off and consider them as finite.
In Figs \ref{fig1}(a) and (b) we show some isodensity contours of the 
dimensionless density $\bar{\rho}_0=\rho_0/(M/L^3)$, equation (\ref{eq_rho0}), 
as functions of $R/L$ and $z/L$ for a `softening parameter' $b/L=0.25$ 
in Fig.\ \ref{fig1}(a) and $b/L=0.75$ in Fig.\ \ref{fig1}(b). Figs \ref{fig2}(a) and (b) 
and Figs \ref{fig3}(a) and (b) display, respectively, isodensity contours of the other dimensionless barred
densities (\ref{eq_rho02}) and (\ref{eq_rho024}) for the same values of 
the parameter $b/L$ as in Figs \ref{fig1}(a) and (b). The softened bars 
retain the same qualitative characteristics as the infinitesimally thin ones, 
e.g.  the isodensity curves in Figs \ref{fig1}(a) and (b) are more elongated 
than those displayed in Figs \ref{fig2}(a) and (b), because the linear density of 
the thin bar (\ref{eq_lambda0}) is less concentrated at its centre than 
the density of the thin bar (\ref{eq_lambda02}).    

\section{Equilibrium points and their stability} \label{sec_eq}

An important aspect related to the morphology of barred galaxies is the 
study of motion of a test particle in the gravitational 
field of a uniform rotating bar. In this section we will discuss the equilibrium 
points and their stability for the motion in the field of the softened bars discussed in Section \ref{sec_bars}. 
For convenience, we place the bar along the $x$-axis, and consider the motion
on the $xy$ plane. For this task, the potentials (\ref{eq_phi0b}), (\ref{eq_phi02b}) 
and (\ref{eq_phi024b}) should be rewritten by replacing $z \rightarrow x$ 
and $R^2 \rightarrow y^2$. In the forthcoming discussion, we shall refer the 
potential-density pair (\ref{eq_phi0b})--(\ref{eq_rho0}) as bar model 1, the pair 
(\ref{eq_phi02b})--(\ref{eq_rho02}) as bar model 2  and the pair (\ref{eq_phi024b})--(\ref{eq_rho024}) 
as bar model 3. 

In a coordinate system attached to the bar that rotates with an (constant) angular 
velocity $\Omega$, the equations of motion of a test particle are 
\begin{align}
\ddot{x}-2\Omega \dot{y} &=-\frac{\partial \Phi_{eff.}}{\partial x} \mbox{,} \label{eq_mot_x}\\
\ddot{y}+2\Omega \dot{x} &=-\frac{\partial \Phi_{eff.}}{\partial y} \mbox{,}  \label{eq_mot_y}
\end{align}
where dots indicate derivatives with respect to time, and $\Phi_{eff.}$ is 
the `effective' potential, 
\begin{equation}
\Phi_{eff.}=\Phi_{bar}-\frac{\Omega^2}{2}\left( x^2+y^2 \right) \mbox{.}
\end{equation}
At an equilibrium point, $\vec{\nabla} \Phi_{eff.}=0$, and the resulting system
of two algebraic equations must be solved to obtain the equilibrium points (Lagrange 
points). Because of the symmetry of the models, the equilibrium points are symmetric with
respect to the $x$- and $y$-axes. One Lagrange point is the origin $(0,0)$, the
pair on the $x$-axis will have coordinates $(\pm x_L,0)$  and the pair on the $y$-axis
will have coordinates $(0,\pm y_L)$. The stability of an equilibrium point is determined
by the linearized equations of motion around it. The following conditions are
necessary and sufficient for an equilibrium point be stable \cite{bt08}: 
\begin{gather} 
\alpha \beta >0 \mbox{,} \label{eq_condI} \\
 -\left( \alpha+\beta+4\Omega^2 \right) <0 \mbox{,} \label{eq_condII} \\
\left( \alpha+\beta +4\Omega^2 \right)^2-4\alpha \beta >0 \mbox{,} \label{eq_condIII}
\end{gather}
and
\begin{equation} \label{eq_partial2}
\alpha= \left( \frac{\partial^2\Phi_{eff.}}{\partial x^2} \right)_{eq.} \text{,} \qquad 
\beta= \left( \frac{\partial^2\Phi_{eff.}}{\partial y^2} \right)_{eq.} \mbox{,}
\end{equation}
where the second derivatives are evaluated at an equilibrium point. We shall calculate 
and analyse the stability of the Lagrange points for each of the three models of bars. 
\subsection{Bar model 1} 

\begin{figure}
\centering
\includegraphics[scale=0.5]{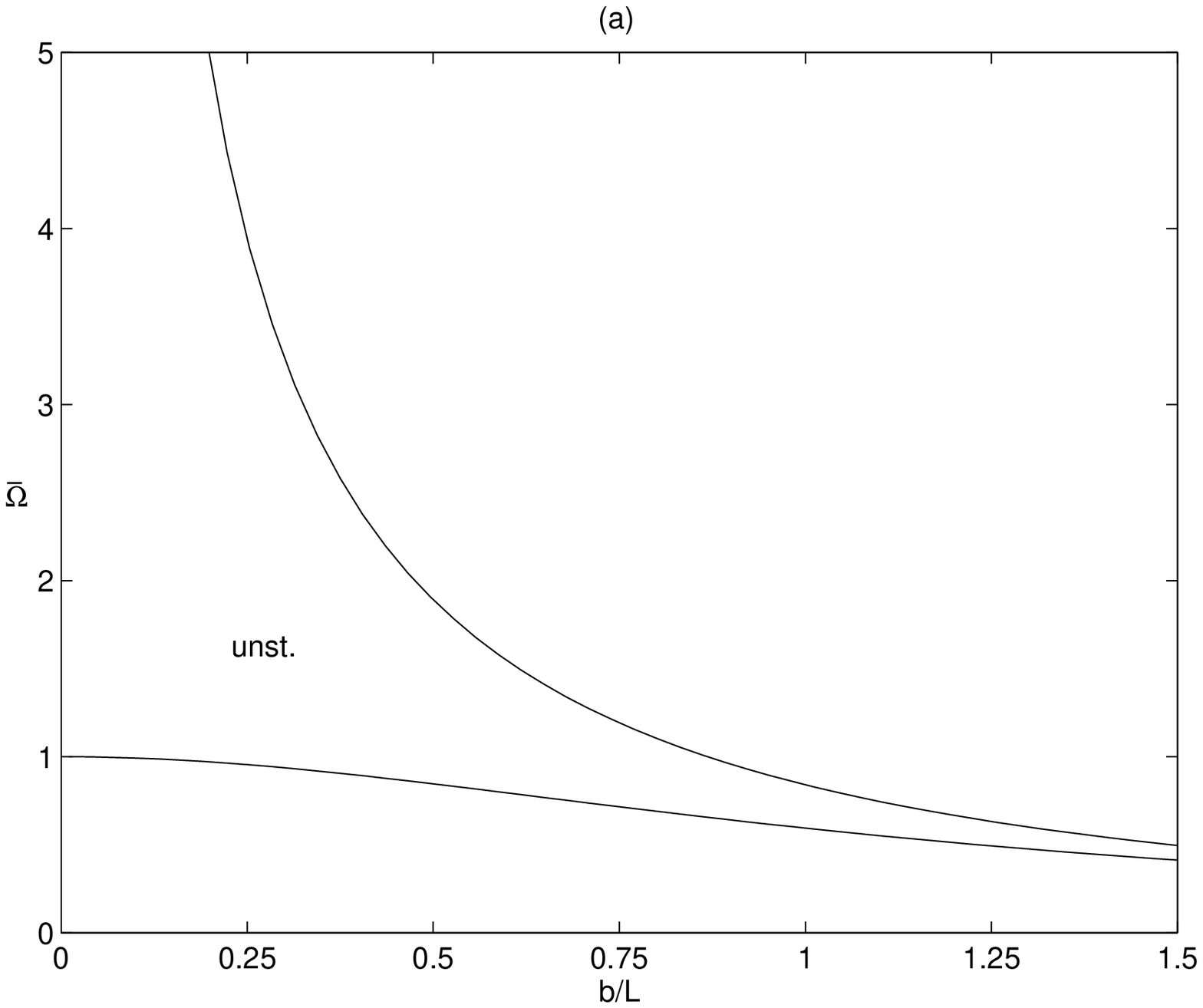}\\
\vspace{0.2cm}
\includegraphics[scale=0.5]{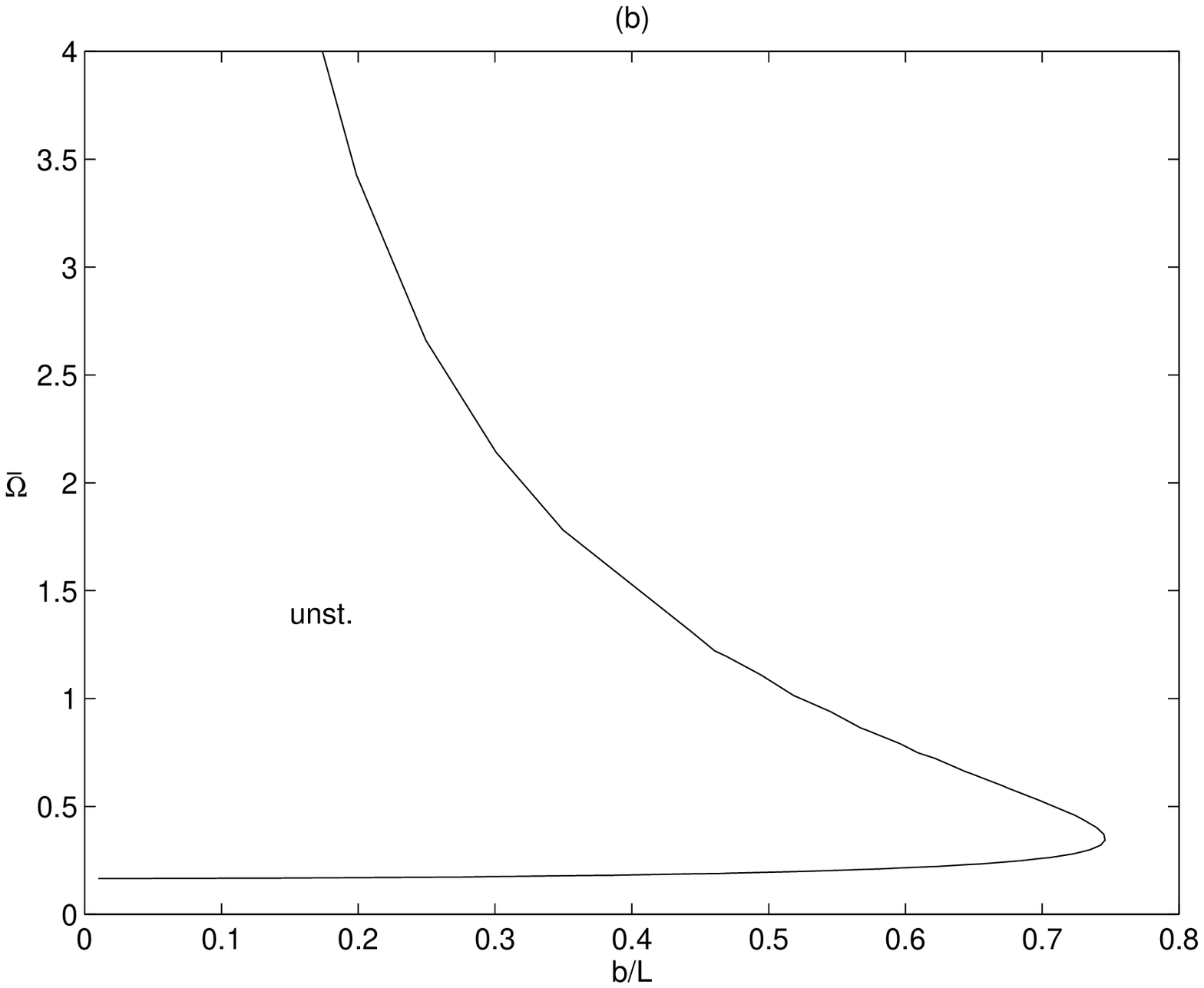}
\caption{(a) Stability diagram of the equilibrium point $(0,0)$ for
bar model 1 as functions of $b/L$ and $\bar{\Omega}=\Omega/\sqrt{GM/L^3}$. 
(b) Stability diagram of the equilibrium point $(0,y_L/L)$
for bar model 1 as functions of $b/L$ and $\bar{\Omega}=\Omega/\sqrt{GM/L^3}$.} \label{fig4}
\end{figure}

At the origin the values of the second derivatives (\ref{eq_partial2}) are 
\begin{equation}
\alpha =\frac{GM}{\left( b^2+L^2\right)^{3/2}}-\Omega^2 \text{,} \qquad
\beta = \frac{GM}{b^2\sqrt{b^2+L^2}}-\Omega^2 \mbox{.}
\end{equation}
One finds analytically that conditions (\ref{eq_condII}) and (\ref{eq_condIII}) are
always satisfied. By condition (\ref{eq_condI}), the origin will be unstable for 
angular velocities in the range 
\begin{equation} \label{eq_omega11}
\sqrt{\frac{GM}{\left( b^2+L^2\right)^{3/2}}} < \Omega < \sqrt{\frac{GM}{b^2\sqrt{b^2+L^2}}} \mbox{.}
\end{equation}
Fig.\ \ref{fig4}(a) shows the stability diagram of the point $(0,0)$,
as functions of $b/L$ and of the dimensionless angular velocity
$\bar{\Omega}=\Omega/\sqrt{GM/L^3}$. The unstable region grows
as the bar becomes more elongated.

On the $y$-axis the equilibrium point is given by the equation 
\begin{equation} \label{eq_omega12}
\frac{GM}{\left( y_L^2+b^2 \right)\sqrt{y_L^2+b^2+L^2}}-\Omega^2=0 \mbox{.}
\end{equation}
In this case the stability is better investigated by a graphical analysis of 
conditions (\ref{eq_condI})--(\ref{eq_condIII}). Fig.\ \ref{fig4}(b) displays 
the stability diagram of the Lagrange points on the $y$-axis as functions of
$b/L$ and $\bar{\Omega}$. For $b/L \gtrapprox 0.75$ the points are stable for
all values of the angular velocity. For lower values of $b/L$ there is an interval
of $\bar{\Omega}$ where the equilibrium points are unstable  and this interval 
becomes larger as $b/L$ approaches zero. In this limit, the points are stable 
for an angular velocity less than 
\begin{equation}
\bar{\Omega}=\frac{13\sqrt{2}-18}{14}\sqrt{5+4\sqrt{2}}\sqrt{2+\sqrt{2}} \approx 0.166 \mbox{.}
\end{equation}

\begin{figure}
\centering
\includegraphics[scale=0.7]{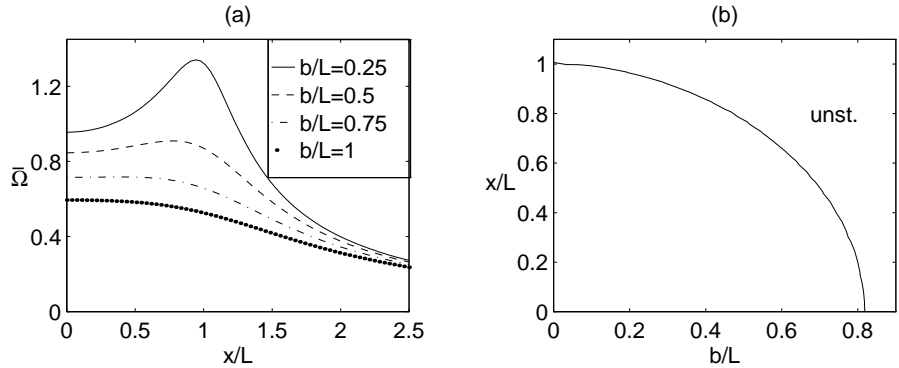}
\caption{(a) Curves of equation (\ref{eq_omega13}) for some 
values of $b/L$. (b) The stability of the equilibrium point(s) $(x_L/L,0)$
for bar model 1 as function of $b/L$.} \label{fig5}
\end{figure}
The equilibrium points on the $x$-axis are given by the equation
\begin{equation} \label{eq_omega13}
\frac{GM\left[ \sqrt{b^2+ \left( x_L+L\right)^2}-\sqrt{b^2+ \left( x_L-L\right)^2} \right]}
{2L\sqrt{b^2+ \left( x_L-L\right)^2}\sqrt{b^2+ \left( x_L+L\right)^2}}-\Omega^2 x_L=0
\end{equation}
For a given value of $\Omega$ the equilibrium point must be found by solving
(\ref{eq_omega13}) numerically. On the other hand, for a given value of $x_L$
one might calculate $\Omega$ directly from (\ref{eq_omega13}).
In Fig.\ \ref{fig5}(a) we plot some curves of $\bar{\Omega}$ as function of 
$x_L/L$ for some values of the parameter $b/L$. It is seen that there may exist
\emph{two} equilibrium points for a given value of the angular velocity, which 
means two pairs of Lagrange points on the $x$-axis. This can happen for values
of $b/L \lessapprox 0.82$. Fig. \ref{fig5}(b) shows the stability of the 
equilibrium point(s) $(x_L/L,0)$ as function of $b/L$. Comparing Figs \ref{fig5}(a)
and (b), we note that when two equilibrium points exist, 
the inner point is always stable, whereas the outer is unstable. When only one 
equilibrium point exists, it is always unstable.
\subsection{Bar model 2}

For this model of bar, the values of the second derivatives (\ref{eq_partial2}) 
at the origin read
\begin{gather} 
\alpha=-\frac{3GM}{2L^3}\ln \left( \frac{\sqrt{b^2+L^2}-L}{\sqrt{b^2+L^2}+L}\right) 
-\frac{3GM}{L^2\sqrt{b^2+L^2}}-\Omega^2 \mbox{,} \notag \\ 
\beta= \frac{3GM}{4L^3}\ln \left( \frac{\sqrt{b^2+L^2}-L}{\sqrt{b^2+L^2}+L}\right)
+\frac{3GM\sqrt{b^2+L^2}}{2b^2L^2}-\Omega^2 \mbox{.}
\end{gather}
The origin will be an unstable equilibrium point for angular velocities in the  
interval 
\begin{multline} \label{eq_omega21} 
\left[ -\frac{3GM}{2L^3}\ln \left( \frac{\sqrt{b^2+L^2}-L}{\sqrt{b^2+L^2}+L}\right)
-\frac{3GM}{L^2\sqrt{b^2+L^2}}\right]^{1/2} < \Omega < \left[ 
\frac{3GM}{4L^3} \right. \\
\left. \times \ln \left( \frac{\sqrt{b^2+L^2}-L}{\sqrt{b^2+L^2}+L}\right)
+\frac{3GM\sqrt{b^2+L^2}}{2b^2L^2}\right]^{1/2} \mbox{.}
\end{multline}
In Fig.\ \ref{fig4}(a) we display the stability diagram of the point $(0,0)$,
as functions of $b/L$ and of the angular velocity
$\bar{\Omega}$. Also in this model the unstable region grows
as the bar becomes more elongated.

\begin{figure}
\centering
\includegraphics[scale=0.5]{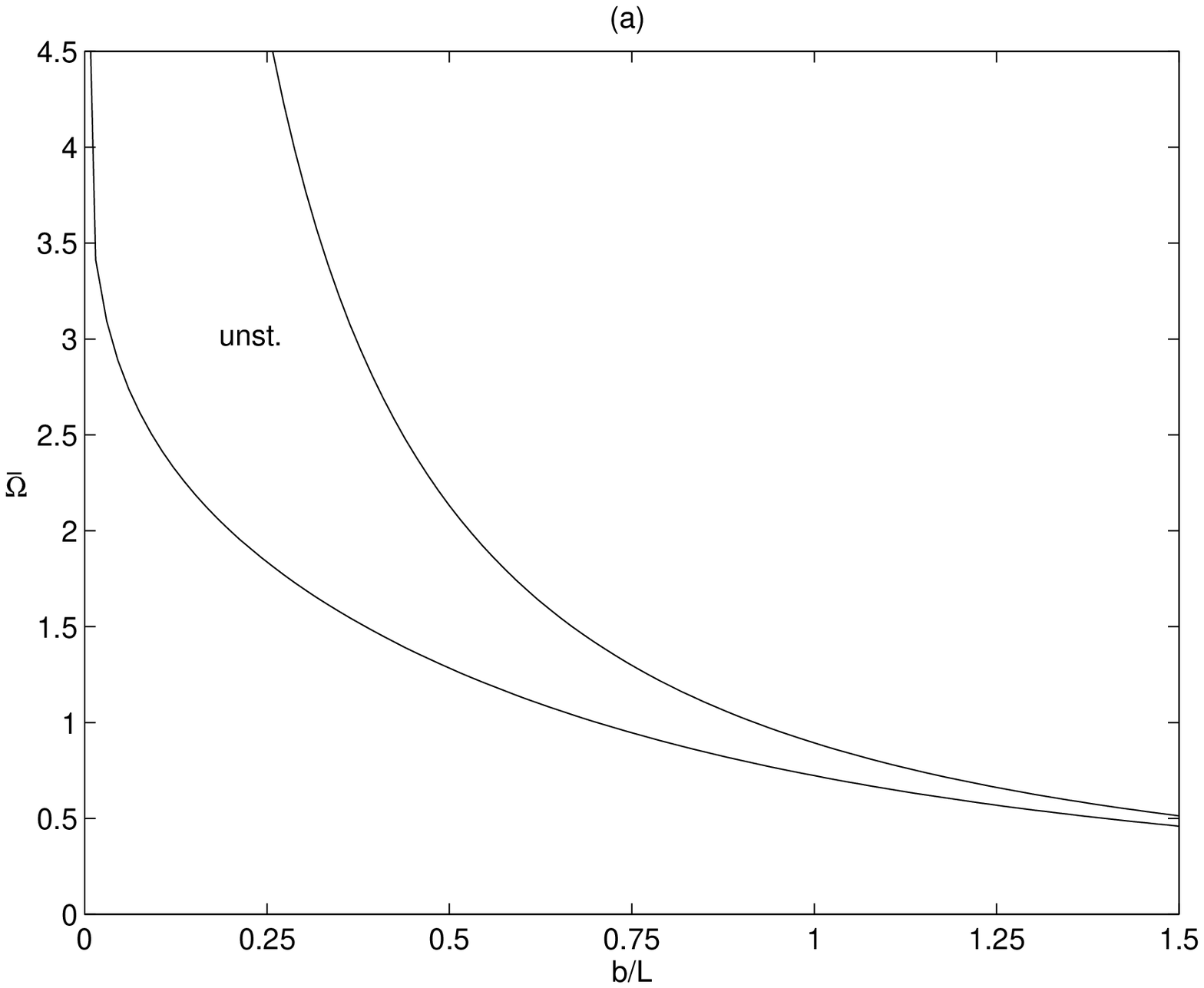}\\ 
\vspace{0.2cm}
\includegraphics[scale=0.5]{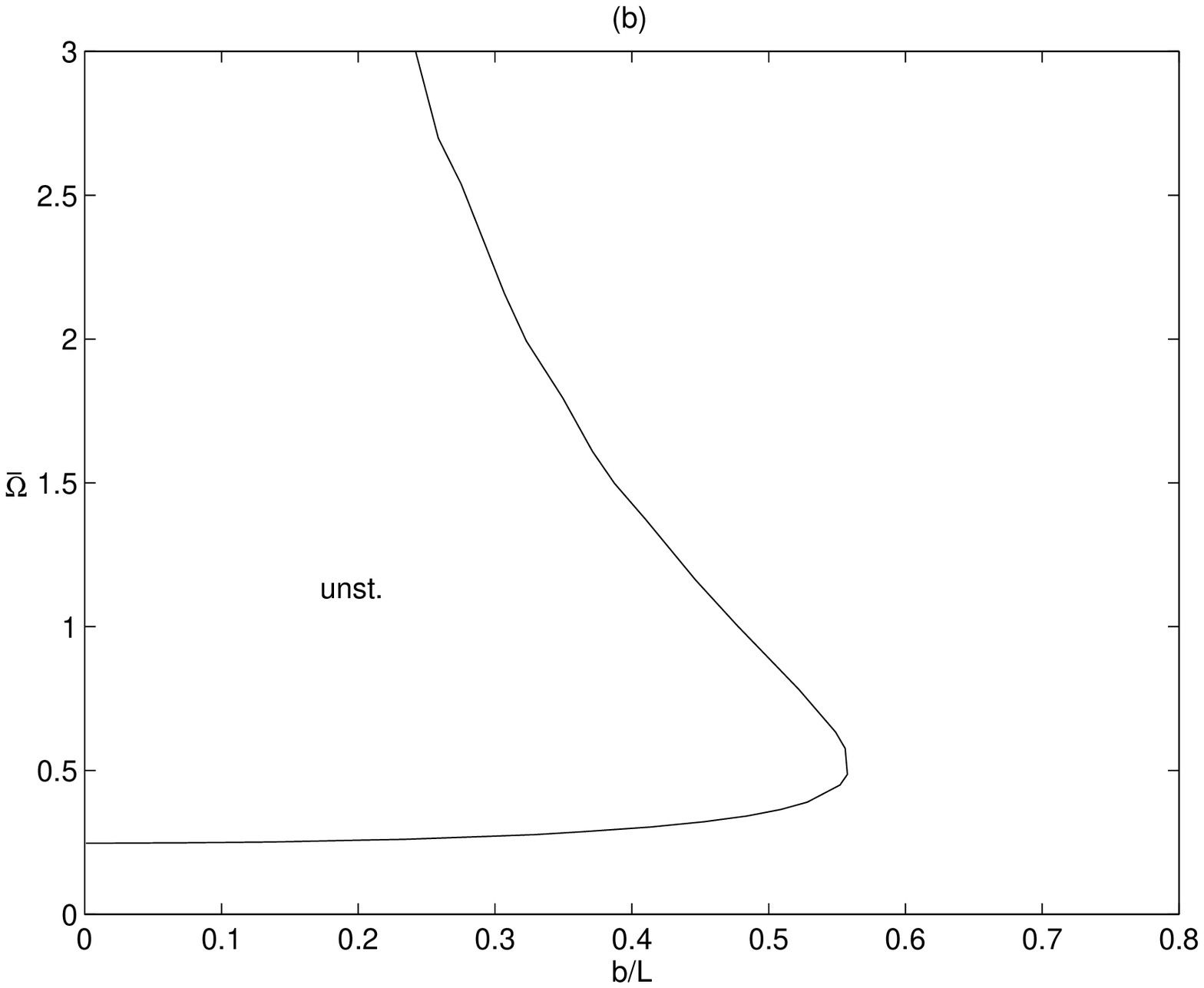}
\caption{(a) Stability diagram of the equilibrium point $(0,0)$ for
bar model 2 as functions of $b/L$ and $\bar{\Omega}=\Omega/\sqrt{GM/L^3}$.
(b) Stability diagram of the equilibrium point $(0,y_L/L)$
for bar model 2 as functions of $b/L$ and $\bar{\Omega}=\Omega/\sqrt{GM/L^3}$.} \label{fig6}
\end{figure}
On the $y$-axis the equilibrium point is given by the equation
\begin{equation} \label{eq_omega22}
\frac{3GM}{4L^3}\ln \left( \frac{\sqrt{y_L^2+b^2+L^2}-L}{\sqrt{y_L^2+b^2+L^2}+L}\right) 
+\frac{3GM\sqrt{y_L^2+b^2+L^2}}{2L^2\left( y_L^2+b^2\right)}-\Omega^2=0 \mbox{.}
\end{equation}
The stability diagram of the Lagrange points on the $y$-axis is 
displayed in Fig.\ \ref{fig6}(b). For $b/L \gtrapprox 0.56$ the points are stable for
all values of the angular velocity. In the limit of infinitesimally thin bar, 
the points are stable for $\bar{\Omega} \lessapprox 0.25$. 

The equilibrium point on the $x$-axis is calculated from
\begin{multline}
-\frac{3GM}{2L^3} \ln \left[ \frac{x_L-L+\sqrt{b^2+\left( x_L-L\right)^2}}
{x_L+L+\sqrt{b^2+\left( x_L+L\right)^2}}\right] \\
-\frac{3GM}{2x_LL^3} \left[ 
\sqrt{b^2+\left( x_L+L\right)^2}-\sqrt{b^2+\left( x_L-L\right)^2}\right] 
-\Omega^2=0 \mbox{.}
\end{multline}
In this case there exists only one equilibrium point for a given value
of the angular velocity, and we found that this point is always unstable.
\subsection{Bar model 3}

\begin{figure}
\centering
\includegraphics[scale=0.7]{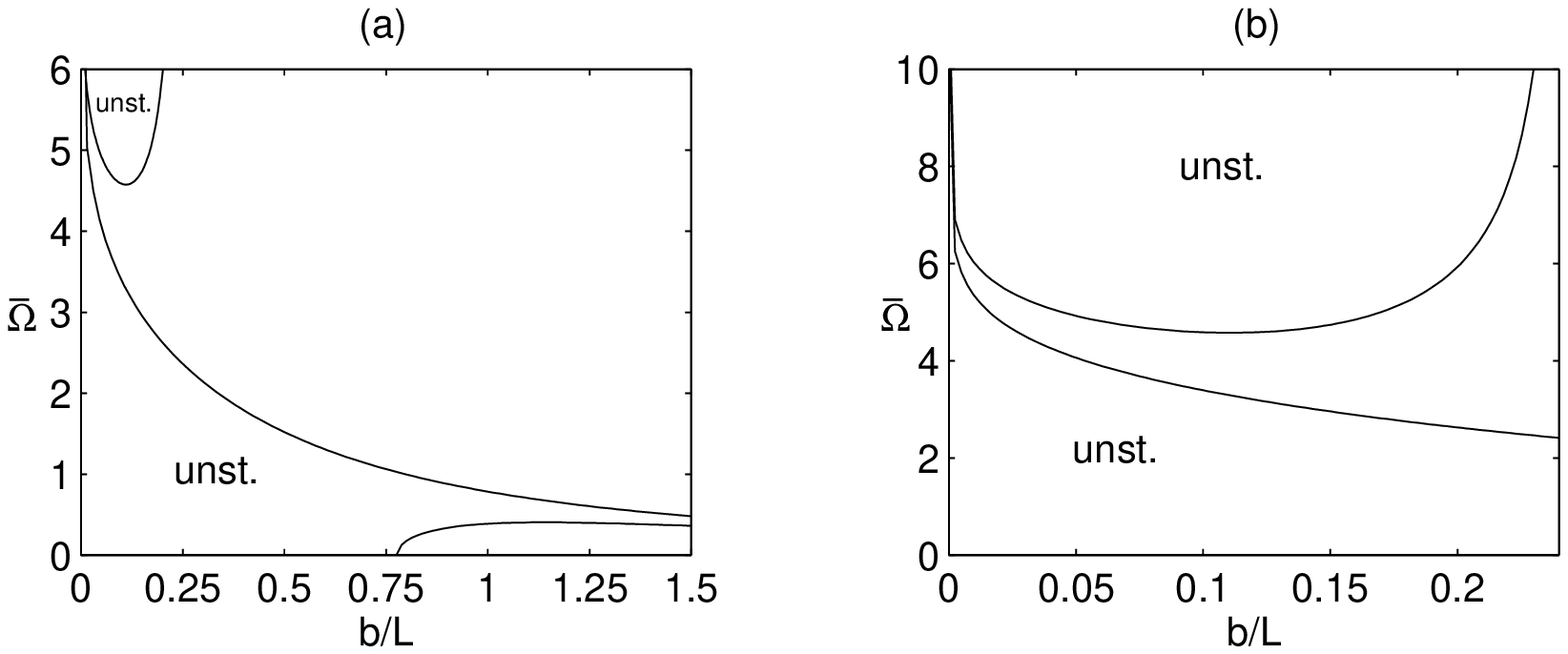}\\
\vspace{0.2cm}
\includegraphics[scale=0.5]{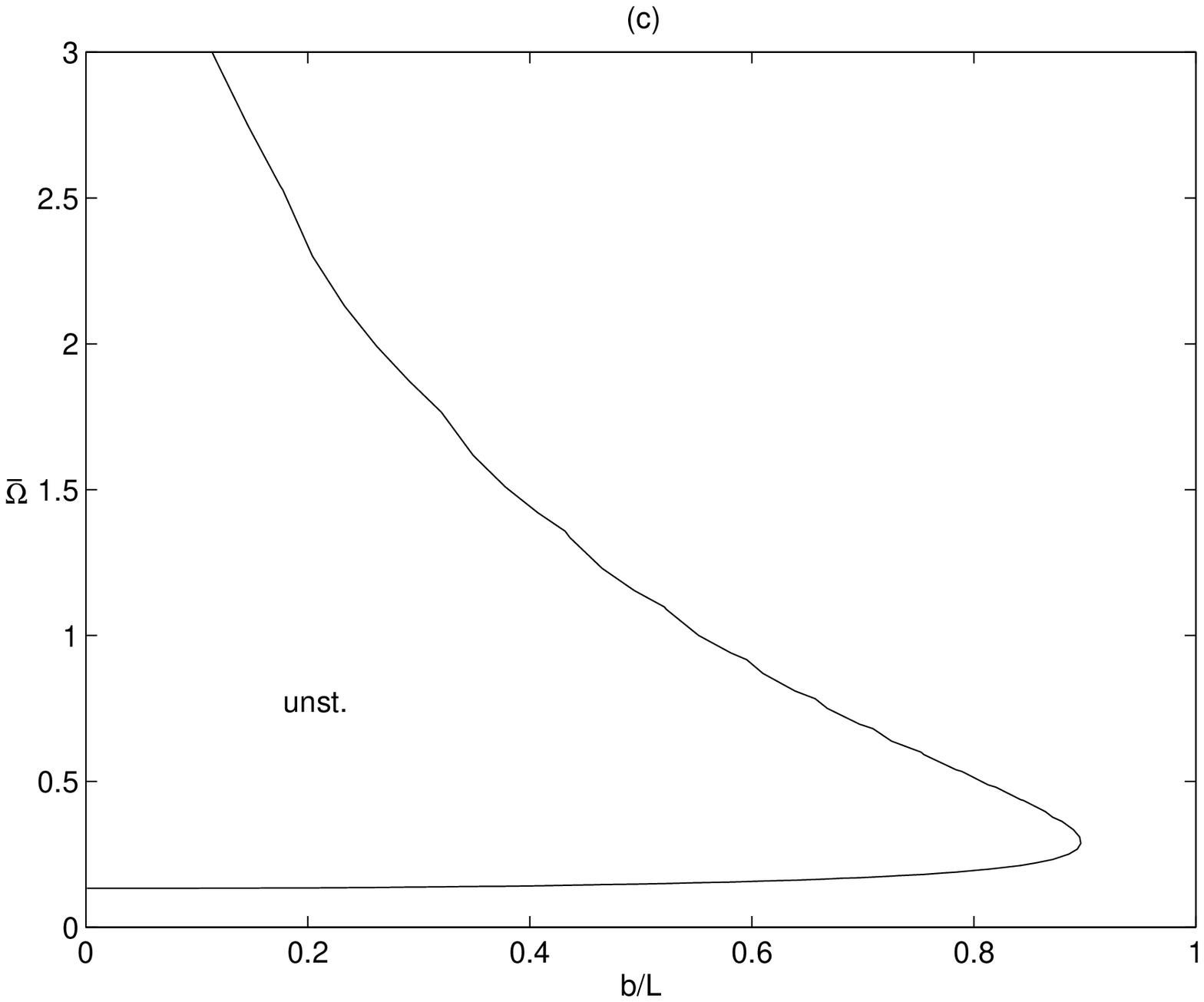}
\caption{(a)--(b) Stability diagram of the equilibrium point $(0,0)$ for
bar model 3 as functions of $b/L$ and $\bar{\Omega}=\Omega/\sqrt{GM/L^3}$.
(c) Stability diagram of the equilibrium point $(0,y_L/L)$
for bar model 3 as functions of $b/L$ and $\bar{\Omega}=\Omega/\sqrt{GM/L^3}$.} \label{fig7}
\end{figure}

For bar model 3, the values of the second derivatives (\ref{eq_partial2}) 
at the origin read
\begin{gather} 
\alpha=\frac{15GM\left( 3b^2+L^2\right)}{2L^5}\ln \left( \frac{\sqrt{b^2+L^2}-L}{\sqrt{b^2+L^2}+L}\right) 
+\frac{15GM\left( 3b^2+2L^2\right)}{L^4\sqrt{b^2+L^2}}-\Omega^2 \mbox{,} \\ 
\beta= -\frac{15GM\left( 3b^2+2L^2\right)}{8L^5}\ln \left( \frac{\sqrt{b^2+L^2}-L}{\sqrt{b^2+L^2}+L}\right)
-\frac{45GM\sqrt{b^2+L^2}}{4L^4}-\Omega^2 \mbox{.}
\end{gather}
The origin will be an unstable equilibrium point for angular velocities in the  
interval $\Omega_1<\Omega<\Omega_2$, where
\begin{gather} \label{eq_omega31} 
\Omega_1= \left[ \frac{15GM\left( 3b^2+L^2\right)}{2L^5}\ln \left( \frac{\sqrt{b^2+L^2}-L}{\sqrt{b^2+L^2}+L}\right) 
+\frac{15GM\left( 3b^2+2L^2\right)}{L^4\sqrt{b^2+L^2}} \right]^{1/2} \mbox{,} \\ 
\Omega_2= \left[ -\frac{15GM\left( 3b^2+2L^2\right)}{8L^5}\ln \left( \frac{\sqrt{b^2+L^2}-L}{\sqrt{b^2+L^2}+L}\right)
-\frac{45GM\sqrt{b^2+L^2}}{4L^4} \right]^{1/2} \mbox{.}
\end{gather}
Furthermore, there is another region of instability for angular velocities greater then $\Omega=\sqrt{A/B}$, 
where
\begin{gather}
A =-\frac{2025G^2M^2\left( 5b^2+2L^2\right)^2}{64L^{10}} \ln^2 \left( \frac{\sqrt{b^2+L^2}-L}{\sqrt{b^2+L^2}+L}\right) \notag \\
-\frac{675G^2M^2\left( 75b^4+85b^2L^2+22L^4\right)}{16L^9\sqrt{b^2+L^2}}  \ln \left( \frac{\sqrt{b^2+L^2}-L}{\sqrt{b^2+L^2}+L}\right) \notag \\
-\frac{225G^2M^2\left( 15b^2+11L^2\right)^2}{16L^8\left( b^2+L^2 \right)} \mbox{,} \\
B =\frac{15GM\left( 9b^2+2L^2 \right)}{L^5} \ln \left( \frac{\sqrt{b^2+L^2}-L}{\sqrt{b^2+L^2}+L}\right)+ 
\frac{30GM\left( 9b^2+5L^2 \right)}{L^4\sqrt{b^2+L^2}} \mbox{.}
\end{gather} 
Figs \ref{fig7}(a) and (b) show the stability diagram of the point $(0,0)$,
as functions of $b/L$ and the angular velocity
$\bar{\Omega}$. Fig.\ \ref{fig7}(b) gives an enlarged view of 
the second region of instability. 

On the $y$-axis the equilibrium point is given by the relation
\begin{multline}
-\frac{15GM\left( 3y_L^2+3b^2+2L^2\right)}{8L^5} \ln \left( \frac{\sqrt{y_L^2+b^2+L^2}-L}{\sqrt{y_L^2+b^2+L^2}+L}\right) \\
-\frac{45GM\sqrt{y_L^2+b^2+L^2}}{4L^4} -\Omega^2=0 \mbox{.}
\end{multline}
The stability diagram of the Lagrange points on the $y$-axis is 
displayed in Fig.\ \ref{fig7}(c). For $b/L \gtrapprox 0.90$ the points are stable for
all values of the angular velocity. In the limit of infinitesimally thin bar, 
the points are stable for $\bar{\Omega} \lessapprox 0.13$. 

\begin{figure}
\centering
\includegraphics[scale=0.7]{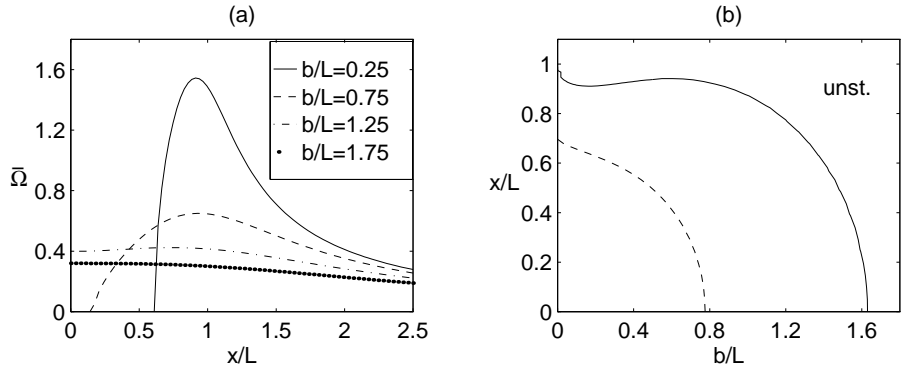}
\caption{(a) Curves of equation (\ref{eq_omega32}) for some 
values of $b/L$. (b) The stability of the equilibrium point(s) $(x_L/L,0)$
for bar model 3 as function of $b/L$. The dashed curve indicates the position 
of the static equilibrium point.} \label{fig8}
\end{figure}

The equilibrium point on the $x$-axis is calculated from
\begin{multline}  \label{eq_omega32}
\frac{15GM\left( 3b^2+L^2-2x_L^2\right)}{2L^5} \ln \left[ \frac{x_L-L+\sqrt{b^2+\left( x_L-L\right)^2}}
{x_L+L+\sqrt{b^2+\left( x_L+L\right)^2}}\right] \\
+\frac{5GM}{2x_LL^5} \left\{ 
\left( 4b^2+L^2-11x_L^2\right)\left[ \sqrt{b^2+\left( x_L+L\right)^2}-\sqrt{b^2+\left( x_L-L\right)^2}\right] \right. \\
\left. +5Lx_L \left[ \sqrt{b^2+\left( x_L+L\right)^2}+\sqrt{b^2+\left( x_L-L\right)^2} \right]\right\} 
-\Omega^2=0 \mbox{.}
\end{multline}
Fig.\ \ref{fig8}(a) shows some curves of $\bar{\Omega}$ as function of 
$x_L/L$ for some values of the parameter $b/L$. As happened with bar model 1, 
 there may also exist two pairs of equilibrium points on the $x$-axis. This is possible 
for values of $b/L \lessapprox 1.63$.  Fig. \ref{fig8}(b) shows the stability of the 
equilibrium point(s) $(x_L/L,0)$ as function of $b/L$. Also here, when two equilibrium points exist, 
the inner point is always stable, whereas the outer is unstable. When only one 
equilibrium point exists, it is always unstable. 

From Fig.\ \ref{fig8}(a) we note a particular feature of this model of bar: even without 
rotation $(\bar{\Omega}=0)$, there is an equilibrium point along the $x$-axis for some values of the parameter
$b/L$ (for instance, $x/L \approx 0.6$ for $b/L=0.25$). This static equilibrium point
exists because the mass density of the bar is not concentrated at the origin (see Fig.\ \ref{fig3}).
In Fig.\ \ref{fig8}(b) the dashed curve indicates the location of this point as function of $b/L$. 
A similar static equilibrium point was found in potential-density pairs for flat rings \cite{vl09}.
\section{Discussion} \label{sec_disc}

We presented analytical potential-density pairs for infinitesimally thin 
and softened bars constructed from an identity that relates multipolar 
solutions of the Einstein equations to Newtonian potentials of bars with 
densities proportional to Legendre polynomials. 
The main advantage of these models is that all
potential density-pairs can be explicitly expressed in terms of 
elementary functions, and bars with a desired density profile can 
be constructed from the set of densities (\ref{eq_lambda}).

As an application of the barred potentials, we calculated 
the equilibrium points for the motion of test particles in 
the gravitational field of three models of rotating bars and analysed their 
stability. The results suggest some conclusions. The equilibrium point $(0,0)$ has the tendency 
to be more stable in bar model 2, and more unstable in bar model 3. 
The stability diagrams for the equilibrium points along the $y$-axis have the
same qualitative behaviour for the three models of bars. On the other hand, 
the properties of the equilibrium points along the $x$-axis seem to be quite 
sensitive to the particular model of bar used. In the case of bar model 2,
the points are always unstable, whereas for bar models 1 and 3, there even
exists the possibility of two pairs of equilibrium points, one being
stable and the other unstable. It is known that the
equilibrium points on the $x$-axis of a homogeneous ellipsoid are always
unstable \cite{d65,m75}. From our three models the bar model 2 has the
nearest shape of an ellipsoid, thus it is not surprising that it exhibits similar
properties. It seems that barred mass distributions with less mass concentrated
around the centre of the bar tend to stabilize the equilibrium points 
along the $x$-axis. Our results are in qualitative agreement with those obtained by 
Michalodimitrakis \cite{m75}, who compared the stability properties of
equilibrium points for a homogeneous ellipsoid and for a homogeneous
parallelepiped.

\section*{Acknowledgments}

DV thanks FAPESP for financial support and PSL thanks FAPESP and CNPq
for partial financial support.

\appendix
\section{Potential-density pairs for softened bars} \label{ap_A}
The expressions for the potentials $\Phi$ and mass densities $\rho$ for the
softened thin bar potentials (\ref{eq_phi0}), (\ref{eq_phi02}) and (\ref{eq_phi024}) 
are given by
\begin{gather} 
\Phi_0= \frac{GM}{2L} \ln \left( \frac{z-L+\mathcal{R}_2}
{z+L+\mathcal{R}_1} \right) \mbox{,} \label{eq_phi0b}\\
\rho_0=\frac{Mb^2}{8\pi L\left( R^2+b^2 \right)^2\mathcal{R}_1^3\mathcal{R}_2^3} 
\left\{ \mathcal{R}_2^3\left( z+L \right) \left[ 3\left( R^2+b^2 \right) +2\left(
z+L \right)^2 \right] \right. \notag \\
\left. - \mathcal{R}_1^3\left( z-L \right) \left[ 3\left( R^2+b^2 \right) 
+2\left( z-L \right)^2 \right] \right\} \mbox{,} \label{eq_rho0}\\
\Phi_{02}= \frac{3GM}{8L^3} \left( R^2+b^2+2L^2-2z^2\right) \ln \left( \frac{z-L+\mathcal{R}_2}
{z+L+\mathcal{R}_1} \right) \notag \\
+ \frac{3GM}{8L^3} \left[ \left( L-3z \right)
\mathcal{R}_1 +\left( L+3z \right) \mathcal{R}_2 \right] \mbox{,} \label{eq_phi02b}\\
\rho_{02}= \frac{3Mb^2}{8\pi L^3\left( R^2+b^2 \right)^2\mathcal{R}_1\mathcal{R}_2}
\left[ \left( R^2+b^2 \right)\left( \mathcal{R}_1-\mathcal{R}_2 \right)z+ 
\mathcal{R}_1\left( z-L \right)^2 \left( z+L \right) \right. \notag \\
\left. -\mathcal{R}_2\left( z+L \right)^2 \left( z-L \right) \right] \mbox{,} \label{eq_rho02} \\
\Phi_{024}=\frac{15GM}{32L^5}\left[ -3\left(R^2+b^2 \right)^2+24\left(R^2+b^2 \right)z^2
-4\left(R^2+b^2 \right)L^2+8z^2L^2 \right. \notag \\
\left. -8z^4 \right] \ln \left( \frac{z-L+\mathcal{R}_2}{z+L+\mathcal{R}_1} \right) 
+ \frac{5GM}{32L^5} \left\{ \left[
-55\left(R^2+b^2 \right)z-9\left(R^2+b^2 \right)L \right. \right. \notag \\
\left. \left. +26z^2L-22zL^2+50z^3-6L^3 \right] \mathcal{R}_2 \right. \notag \\
\left. +\left[ 55\left(R^2+b^2 \right)z-9\left(R^2+b^2 \right)L+26z^2L+22zL^2-50z^3-6L^3 \right]
\mathcal{R}_1 \right\} \label{eq_phi024b}\mbox{,} \\
\rho_{024}=\frac{45Mb^2}{16 \pi L^5} \ln \left( \frac{z-L+\mathcal{R}_2}
{z+L+\mathcal{R}_1} \right)+\frac{15Mb^2}{16\pi L^5\left( R^2+b^2 \right)^2\mathcal{R}_1\mathcal{R}_2} \notag \\
\times \left\{ \mathcal{R}_1 \left[ \left( R^2+b^2 \right)^2\left( 5z+3L \right)+
\left( R^2+b^2 \right) \left( 7z^3-5z^2L-zL^2+L^3 \right) \right. \right. \notag \\
\left. \left. +2z^2\left( z+L \right)\left( z-L \right)^2 
\right]+\mathcal{R}_2 \left[ \left( R^2+b^2 \right)^2\left( -5z+3L \right)+
\left( R^2+b^2 \right) \right. \right. \notag \\
\left. \left. \times \left( -7z^3 -5z^2L+zL^2+L^3 \right)-2z^2\left( z-L \right)\left( z+L \right)^2
\right] \right\} \mbox{,} \label{eq_rho024}
\end{gather}
where $\mathcal{R}_1=\sqrt{R^2+b^2+\left( z+L \right)^2}$ and $\mathcal{R}_2=\sqrt{R^2+b^2+\left( z-L \right)^2}$.
\end{document}